\documentstyle[twocolumn,prl,floats,aps]{revtex}

\def\bea{\begin{eqnarray}}
\def\eea{\end{eqnarray}}
\def\ben{\begin{equation}}
\def\een{\end{equation}}
\def\benu{\begin{enumerate}}
\def\enu{\end{enumerate}}

\def\n{n}

\def\sss{\scriptscriptstyle\rm}

\def\1var{(\bx_1...\bx\N)}

\def\br{{\bf r}}

\def\bx{{x}}

\def\bj{{\bf j}}

\def\s{_{\sss S}}
\def\xc{_{\sss XC}}

\def\N{_{\sss N}}

\def\ext{_{\rm ext}}

\def\unif{^{\rm unif}}

\def\ee{_{\rm ee}}

\def\ALDA{^{\rm ALDA}}

\def\td{time-dependent~}

\def\sph_int{ {\int d^3 r}}

\def\PRL{Phys. Rev. Letts.\ }
\def\JCP{J. Chem. Phys.\ }

\input psfig.sty
\input rotate

\begin{document}
\headheight 50pt

\title{Initial-state dependence
in time-dependent density functional theory}
\author{
N.T. Maitra and K.  Burke \\ Departments
of Chemistry and Physics, Rutgers University,
\\Piscataway, NJ 08854, USA\\{\it submitted to Phys. Rev. A.}\\}
\maketitle

\begin{abstract}
Time-dependent density functionals in principle depend on the
initial state of a system, but
this is ignored in functional approximations
presently in use. 
For one electron, it is shown that there is no
initial-state dependence:  for any density, only one initial
state produces a well-behaved potential.
For two non-interacting electrons with the
same spin in one-dimension,
an initial potential that makes an alternative initial wavefunction
evolve with the same density and current as a ground state
is calculated.  This potential is well-behaved, and can be made
arbitrarily different from the original potential.

\end{abstract}
\section{Introduction and conclusions}
\label{s:intro}
Ground-state density functional theory\cite{HK64,KS65}
has had an enormous impact on solid-state physics since its invention, and
on
quantum chemistry in recent years \cite{Kb99}.
Time-dependent density
functional theory (TDDFT) allows the external potential
acting on the electrons to be time-dependent, and so
opens the door to a wealth of interesting and
important phenomena that are not easily accessible, if at all, within
the static theory.
Important examples include atomic and molecular collisions\cite{YB00},
atoms and molecules in intense laser fields\cite{TC98},
electronic transition energies and oscillator strengths\cite{YB99,YBb99},
frequency-dependent polarizabilities and hyperpolarizabilities,
etc.\cite{GSB95}, and there has been an explosion of time-dependent Kohn-Sham
calculations in all these fields.   In almost all these calculations,
the ubiquitous adiabatic local density approximation (ALDA)\cite{ZS80,GDP96} 
is used to
approximate the unknown time-dependent exchange-correlation potential,
i.e., $v\xc\ALDA[\n](\br t) = v\xc\unif(\n(\br t))$, where $v\xc\unif(\n)$
is the {\em ground-state} exchange-correlation potential of a uniform
electron gas of density $\n$.   While this seems adequate for many
purposes\cite{MSb90}, little is known about its accuracy under the myriad
of circumstances in which it has been applied.

Runge and Gross\cite{RG84} formally established time-dependent density
functional theory (TDDFT),
showing that
for a given initial state,
the evolving density uniquely identifies the (time-dependent)
potential.  This established the correspondence of a unique
non-interacting system to each interacting system and so a set of
one-particle Kohn-Sham equations, much like in the static theory. This
one-to-one mapping between densities and potentials is the
time-dependent analog of the Hohenberg-Kohn theorem, but with a major
difference: in the time-dependent case, the mapping is unique only for
a specified initial state. The functionals in TDDFT depend not only on
the time-dependent density but also on the initial state. This
dependence is largely unexplored and indeed often neglected, for example
in the ALDA for the exchange-correlation potential mentioned above.

What do we mean by an initial-state dependence?
In the ground-state theory, there is a simple one-to-one
relation between ground-state densities
and Kohn-Sham potentials $v\s$, assuming they exist.   For example, for
one electron in one dimension, we can easily invert the Schr{\"o}dinger
equation, to yield
\ben
v\s(x) = \frac{d^2 
{\sqrt{\n(x)}} / d x^2}
{2 {\sqrt{\n(x)}} }
+\epsilon
\label{vsone}
\een
where $\n(x)$ is the ground-state density. We use atomic units 
throughout ($\hbar = m = e^2 =1$).  For $N$ electrons in three dimensions,
one can easily imagine continuously altering $v\s(\br)$, the Kohn-Sham potential,
solving the Schr{\"o}dinger equation, finding the orbitals
and calculating their density, until the correct $v\s(\br)$
is found to reproduce the desired density.  By the Hohenberg-Kohn
theorem, this potential is unique, and several clever
schemes for implementing this idea appear in the literature
\cite{CS91,ZP92,G92,N93,WP93,UG93,LLMK93,LB94}.
This procedure could in principle be implemented for  interacting
electrons, if a sufficiently versatile
and accurate interacting Schr{\"o}dinger equation-solver
were available.

Now consider the one-dimensional one-electron density
\ben
\n(x)=2 x^2\ \exp(-x^2)/{\sqrt{\pi}},
\label{none}
\een
(actually the density of the first excited state of a harmonic
oscillator) .  If we consider this as a ground-state
density, we are in for an unpleasant surprise.   Feeding it
into Eq. (\ref{vsone}), we find that
the potential which generates this density  is parabolic
almost everywhere ($x^2/2$), but has a nasty unphysical spike
at $x=0$, of the form $\delta(x)/|x|$.   We usually exclude
such potentials from consideration\cite{DG90}, and regard this density
as not being $v$-representable.

But now imagine that density as being the density of
a first excited state.   In that case, the relation
between density and potential is {\em different}, because
the orbital changes sign at the node.  The mapping becomes
\ben
v\s(x) = \frac{d^2 \left( {\rm sgn}(x-x_0)
{\sqrt{\n(x)}} \right)/ d x^2}
{2 \left( {\rm sgn}(x-x_0){\sqrt{\n(x)}} \right)}
+\epsilon
\label{Vstwo}
\een
where ${\rm sgn}(x) = 1$ for $x>0$ and -1 for $x<0$,
and $n(x_0)=0$.  If we use
this mapping, we find a perfectly smooth parabolic well ($x^2/2)$.
This is a simple example of how the {\em mapping} between densities
and potentials depends on the initial state.   

More generally,
for any given time-dependent density $\n(\br t)$, we
ask how the potential $v(\br t)$ whose wavefunction yields that
density depends on the choice of initial wavefunction $\Psi_0$, i.e.,
in general $v[\Psi_0,\n](\br t)$.  Our aim in this paper is to
explicitly calculate two different potentials giving rise to
the same time-dependent density by having two different initial
states.
Note that even finding such a case is non-trivial.
The choice of wavefunctions is
greatly restricted by the time-dependent density.
As van Leeuwen has pointed out \cite{L99}, the continuity
equation $\dot n = -\nabla \cdot {\bf j}$  implies that 
 only
wavefunctions that have the correct initial current
are candidates for generating a given time-dependent
density.  
Van Leeuwen  also shows how
to explicitly construct the potential generating a given
density from an allowed initial wavefunction using equations
of motion.

Why is this important?  The exchange-correlation
potential, $v\xc(\br t)$, of TDDFT is the difference between a Kohn-Sham
potential and the sum of the external
and Hartree potentials.  Since both the interacting and
non-interacting mappings can depend on the choice of initial
state, this potential is a functional of both initial
states and the density, i.e., $v\xc[\n,\Psi_0,\Phi_0] (\br t)$.
But in common practice, only the dependence on the density is
approximated.  We show below that this misses significant
dependences on the initial state, which can in turn be
related to memory effects, i.e., dependences on the density
at prior times.

\begin{figure}
\centerline{
\psfig{figure=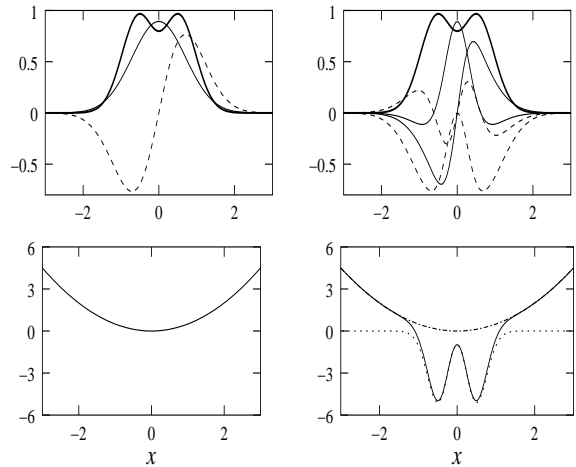,width=8cm,height=6.7cm}}
\label{fig:eg2}
\caption{The top left-hand plot shows the ground-state
orbitals $\phi_1$ (solid)
and $\phi_2$ (dashed) and
their density $n$ (thick solid line) for the harmonic
potential in the lower left-hand plot.
The top right-hand plot contains the real and imaginary parts of 
alternative orbitals $\tilde\phi_1$(solid) and $\tilde\phi_2$(dashed), 
and their density, $n$(thick solid), while 
below is the unique initial potential $\tilde v$ (solid line) that
keeps the density constant.}
\end{figure}
In the special case of one electron, we prove in section \ref{s:one}
that only one initial state has a physically well-behaved potential.
Any attempt to find
another initial state which evolves in a different potential with the
same evolving density, results in a ``pathological'' potential.
The potential either has the strong features at nodes mentioned
above, or 
rapidly plunges to minus infinity at large
distances where the density decays.
(How such a potential can support a localized density is
discussed in section \ref{s:examples}.)
Such non-physical states and potentials are excluded from
consideration (as indeed they are in the Runge-Gross
theorem).
Thus there is no
initial-state dependence for one electron.

We might then reasonably ask, can we ever find a well-behaved
potential for more than one allowed initial wavefunction?
The answer is yes, which we demonstrate with a specific example.
Consider two non-interacting electrons of
the same spin in a harmonic well.  In the ground state of this
two electron system, the first electron occupies the oscillator ground
state, and the second occupies the first excited state, as shown in
Fig. \ref{fig:eg2}.  If we keep the potential constant, the density will not
change.
By multiplying each orbital by a spatially varying phase, 
and choosing these phases to make the current vanish, we 
find an allowed alternative initial state
(see section
\ref{s:examples} for details).
Van Leeuwen's prescription
then yields the unique potential which makes this
wavefunction evolve with the same density.
The difference
is perfectly well-behaved, and can be made arbitrarily large
by adjusting a constant in the phases of the alternative orbitals.
To our knowledge,
this is the first explicit construction of two different potentials
that yield the same time-dependent density.
Other examples are given in section
\ref{s:examples}.

Now imagine the density of Fig. \ref{fig:eg2} were the
ground-state density of some {\em interacting} two-electron system, in
some external potential $v\ext(x)$.   Then both potentials shown in
lower panels of Fig. \ref{fig:eg2}
are possible Kohn-Sham potentials, $v\s(x)$, for this system.  
Since the
Hartree potential is uniquely determined by the density,
we have two very different
exchange-correlation potentials, differing by the amount shown.
In fact, different choices of initial wavefunction allow us to
make the two dips arbitrarily deep or small.
Any
purely density functional approximation misses this effect entirely,
and will produce the same exchange-correlation potential for all cases.

So, even in the simplest case of non-degenerate
interacting and Kohn-Sham ground states, one can choose
an alternative Kohn-Sham initial state, whose potential
will look very different from that which evolves from 
the initial ground state.   
In practice, 
the majority of applications of TDDFT presently involve
response properties of the ground-state of a system,
and one naturally choses to start the Kohn-Sham system
in its ground state.  This choice is also dictated
by the common use of adiabatic approximations for exchange-correlation
potentials, which are approximate {\em ground-state} potentials
evaluated on the instantaneous density.   Such models will
clearly be inaccurate even at $t=0$ if we start our
Kohn-Sham calculation in any state other than its ground state.

The initial-state dependence of functionals is deeply connected to the
issue of memory effects which are ignored in most TDDFT functional
approximations used today. Yet these can often play a large role in
exchange-correlation energies in fully time-dependent
(i.e. non-perturbative) calculations\cite{HMB00} as well as giving
rise to frequency-dependence of the exchange-correlation kernel in
linear response theory ($f\xc(\omega)$) \cite{LGP00}.  Functionals in
general depend not only on the density at the present time, but also
on its history. They may have a very non-local (in time) dependence on
the density.  But still more about the past is required: the
functional is also haunted by the initial wavefunction. The initial
state dependence is inextricably linked to the history of the density
and in fact can be absorbed into density-dependence along a pseudo
pre-history \cite{MBc00}. The results of this current paper shed some
light on the importance of memory effects arising from the initial
wavefunction.

To summarize, we have shown that there is no initial-state
dependence for one electron, and
that there can be arbitrarily large
initial-state dependence for two electrons.

\section{Theory}
\label{s:theory}
Consider a density $n(\br t)$ evolving in time under a
(time-dependent) potential $v(\br t)$. Can we obtain the same evolving
density $n(\br t)$ by propagating some different initial state in a
different potential $\tilde v(\br t)$? This was answered in the
affirmative in Ref. \cite{L99} under the condition that the two initial
states have the same initial density and initial first time-derivative
of the density.  Here we shall show that additional restrictions are
required on the initial state for this statement to hold. In the
one-electron case, the additional restrictions are so strong that
there is {\it no} other initial state that evolves with the same
density as another does in a different potential.

\subsection{One electron}
\label{s:one}
Any two
one-electron wavefunctions $\phi(\br t)$ and
$\tilde\phi(\br t)$ with the
same density $n(\br t) = \vert \phi(\br t) \vert^2
=  \vert \tilde\phi(\br t) \vert^2$, 
are related by a spatial and time dependent phase factor:
\ben \tilde\phi(\br t) = \phi(\br t)\ \exp\left(i\alpha(\br t)\right)
\label{eq:psi}
\een
where the phase $\alpha(\br t)$ is real.
The evolution of each wavefunction is determined by the time-dependent
Schr\"odinger equation with its potential (dot implies
a time derivative):
\ben
\left(-\nabla^2/2 +v(\br  t)\right)\phi(\br t)
 = i\dot \phi(\br t).
\label{eq:TDSE}
\een
Both will satisfy the continuity equation:
\ben \dot \n (\br t) = -\nabla
\cdot {\bf j} (\br t)
\label{eq:continuity}
\een
where the current density of a wavefunction $\phi$ is  
\ben
{\bf j} (\br t) = i\left(\phi (\br t)\nabla \phi^* (\br t)
- \phi^*(\br t)\nabla \phi(\br t)\right)/2.
\label{eq:current}
\een
Substituting $\tilde\phi(\br t)$ from Eq. (\ref{eq:psi}) 
into Eq. (\ref{eq:current}), we obtain
\ben
\Delta {\bf j}(\br t) = \n(\br t) \nabla \alpha(\br t).
\label{delj}
\een
(We use the notation $\Delta a$ to denote $\tilde a - a$.)
Because the densities are the same for all times, $\Delta \dot n(\br t)=0$,
so by Eq. (\ref{eq:continuity})
\ben
\nabla \cdot [n(\br t) \nabla \alpha(\br t)]=0
\label{eq:divngradalpha}
\een 
Integrating
Eq. (\ref{eq:divngradalpha}) with $\alpha(\br t)$ and performing the
integral by parts, we find
\ben
\int d^3r\ n(\br t) \vert \nabla \alpha(\br t)\vert^2 =0.
\label{eq:intbyparts}
\een 
We have taken the surface term $\int d^2{\bf S}\cdot (\alpha n
\nabla\alpha)$ evaluated on a closed surface at infinity, to be zero:
this arises from the physical requirement that at infinity, where the
electron density decays, any physical potential remains finite. 
(In fact this condition is required in
the proof of the Runge-Gross theorem \cite{RG84}).
If the surface term did not
vanish, then $\alpha\nabla\alpha$ must grow at least as fast as
$(r^2n(\bf r))^{-1}$ as $r$ approaches infinity. This would lead to a
potential that slides down to $-\infty$ which can be seen by inversion
of the time-dependent Schr\"odinger equation in the limit of large
distances. The state would oscillate infinitely wildly at large
distances in the tails of the density but the decay of the density is
not enough to compensate for the energy that the wild oscillations
impart: this state would have infinite kinetic energy, momentum and
potential energy.  (We shall see this explicitly in section 
\ref{s:one-e in one-d}). 
So for physical situations, the surface term vanishes.

Because the integrand above cannot be negative, yet it integrates to
zero, the integrand itself must be zero everywhere.  Thus $\nabla
\alpha(\br t) = 0$ everywhere except perhaps at nodes of the
wavefunction where $n(\br_0 t) = 0$. In fact, even at the nodes,
$\nabla \alpha(\br_0 t)= 0$ to avoid highly singular potentials: if
$\nabla\alpha$ was finite at the nodes and zero everywhere else, then
as a distribution, it is equivalent to being zero everywhere, for
example its integral $\alpha$ is constant. There remains the
possibility that $\nabla\alpha$ is a sum of delta functions centered
at the nodes, however this leads to potentials which are highly
singular at the nodes, as in the introduction.
Such unphysical
potentials are excluded from consideration, so that
$\nabla\alpha(\br t)=0$, or
$\alpha(\br t) = c(t)$.
The wavefunctions $\phi(\br t)$ and $\tilde\phi(\br t)$ can therefore 
differ only by an irrelevant
\td phase.
In particular, this means that only one initial state and one
potential can give rise to a particular density, i.e, the evolving
density is enough to completely determine the potential and
initial state.

The one-electron case is a simple counter-example to the conclusions
in Ref. \cite{L99}, which rely
on the existence of
a solution to 
\ben
\nabla \cdot [n\nabla \Delta v] = \eta({\bf r}  t),
\label{eq:SL}
\een
where $\eta({\bf r}  t)$ involves expectation values of derivatives of
the momentum-stress tensor and derivatives of the interaction
(see section \ref{s:many}).
This
is to be solved for the potential subject to the requirements
that the two initial states have the same 
$n(\br 0)$ and $\dot n(\br 0)$, and 
that $\nabla\Delta v\to 0$ as
$r\to\infty$.  
The  two initial wavefunctions in the one-electron case 
have the same initial $n(\br 0)$ (Eq. \ref{eq:psi}) and $\dot n(\br 0)$ (Eq. \ref{eq:divngradalpha}) but no two
physical potentials exist under which they would evolve with the same
density, because there is no
solution to Eq. (\ref{eq:SL}) subject to the boundary condition
that $\nabla\Delta v = 0$. (For an explicit demonstration of this in
one-dimension, see section \ref{s:one-e in one-d}.) We shall come back to this point at the end of section \ref{s:many}.

\subsection{The many-electron case}
\label{s:many}

In this section we follow van Leeuwen's prescription to
find the potential needed to make a given initial state evolve
with the same density as that of another.  However we simplify the
equations there somewhat to make the search for the solution of the
potential easier. Given an initial state $\Psi_0$ which
evolves with density $n({\bf r} t)$ in a potential $v({\bf
r} t)$, we solve for the potential $\tilde v({\bf r}  t)$ in which a
state $\tilde\Psi$ evolves with the same density $n({\bf
r} t)$. If we require $\tilde\Psi$ to have the same initial density and
initial first time-derivative of the density, then a solution for
$\tilde v$ may be obtained from equating the equations for the second
derivatives of the density for each wavefunction,
subject to an appropriate boundary condition like $\Delta v \to
0$ at large distances.  
We are not guaranteed that such a solution exists:
the wavefunction must have the additional restriction that the initial
potential computed in this way is bounded at infinity.

The equation of motion for $\dot n$ yields (Eq. (15) of Ref. \cite{L99}):
\ben \ddot n (\br t) =\nabla\cdot\left[n(\br t)\nabla v(\br t) 
+{\bf t}(\br t)+{\bf f\ee}(\br t)
\right]
\label{eq:secondderiv2}
\een
where 
\ben
{\bf t} (\br t)
=(\nabla'-\nabla)(\nabla^2 -\nabla'^2)
\rho_1(\br'\br t)\vert_{\br'=\br}/4
\label{eq:tildet}
\een
and 
\ben
{\bf f\ee} (\br t) = \int d^3r' P(\br'\br t) \nabla v\ee(|\br'-\br|)/2
\label{eq:tildew}
\een
where $\rho_1(\br'\br t)$
is the (off-diagonal)
one-electron reduced density matrix, $P(\br'\br t)$
is the pair
density (diagonal two-electron reduced density matrix)
and $v\ee(u)$ is the two-particle interaction, {\it e.g.} $1/u$.
Here and following, $\nabla$ and $\nabla'$ indicate the partial gradient
operators with respect to $\br$ and $\br'$ respectively. In
Eq. (\ref{eq:tildet}) and similar following equations, $\br'$ is set
equal to $\br$ after the derivatives are taken.

The idea \cite{L99} is to subtract Eq. (\ref{eq:secondderiv2}) for
wavefunction $\Psi$ from that for wavefunction $\tilde \Psi$, and
require that $\ddot n$ is the same for each. First, we simplify
the kinetic-type term, ${\bf t}$.
Differentiating the continuity equation (Eq. (\ref{eq:continuity})) implies
\ben
(\nabla+\nabla') (\nabla^2-\nabla'^2) \Delta
\rho_1(\br'\br t)\vert_{\br'=\br}
=0
\een 
This equality enables us to incorporate the satisfaction of the
equation of continuity in Eq.
(\ref{eq:secondderiv2})  (when we subtract the equation for $\Psi$ from
that for $\tilde \Psi$) and it also simplifies the kinetic-type term:
\ben
\Delta {\bf t} (\br t)=
-\nabla \left(\nabla^2-\nabla'^2\right)
\Delta \rho_1(\br'\br t)\vert_{\br'=\br}/2.
\label{eq:tildet2}
\een 
(We note that although this is no longer explicitly real, it is in
fact real for states with the same density and first time-derivative.) 
So our simplified equation to solve becomes 
\ben
\nabla \cdot \left[n\nabla\Delta v +\Delta {\bf t}+ \Delta {\bf f}\ee\right]
=0
\label{eq:simpSL}
\een
where $\Delta {\bf t}$ is given by Eq. (\ref{eq:tildet2})
and $\Delta {\bf f}\ee$ is
given by Eq. (\ref{eq:tildew}), applied to the pair density difference.

To calculate the derivatives in the kinetic-type term,
we define
\ben
\gamma(\br\br' t) = \Delta \log\rho_1(\br' \br t)
\label{eq:rho1}
\een 
Note that $\gamma$ vanishes at $\br=\br'$, and
since $\rho_1(\br' \br t) = \rho_1^*(\br \br' t)$,
$\gamma(\br'\br t) = \gamma^* (\br\br' t)$.
These relations also imply that
$\nabla^m\gamma(\br'\br t)\vert_{\br'=\br}
=\nabla^m\gamma^*(\br\br' t)\vert_{\br'=\br}$.
Writing
\ben
\gamma(\br\br' t) = \beta(\br\br' t) + i\alpha(\br\br' t)
\label{alphadef}
\een
where $\alpha$ and $\beta$ are real functions, we also
find
$\nabla\beta(\br' \br t)\vert_{\br'=\br} =0$,
since
$\nabla_\br[\beta(\br' \br t)\vert_{\br'=\br}]=0$.  
Also,
$\nabla \cdot \nabla' \alpha(\br' \br t)\vert_{\br'=\br} =0$,
which follows from the antisymmetry of $\alpha$.
The generalization of Eq. (\ref{delj}) is
\ben
\Delta \bj (\br t)
=\n(\br t)\nabla\alpha(\br' \br t)\vert_{\br'=\br}
\een
Continuity  (Eq.(\ref{eq:continuity})) then gives us a condition on
the near-diagonal elements of $\alpha$:
\ben
\nabla\cdot\left[n(\br t)\nabla\alpha(\br' \br t)\vert_{\br'=\br}\right]=0
\label{eq:divngradalpdiag}
\een
Using all these results in Eq. (\ref{eq:simpSL}), we find
\bea
\nonumber
\nabla \Delta v(\br t)&=&\left(\dot n\nabla \alpha -
\nabla^2\alpha\nabla \Im\rho_1 +\nabla \times {\bf B}\right)/n\\
\nonumber
&+&\nabla\Re\left(\nabla\gamma\cdot\nabla\rho_1-
\nabla'\gamma\cdot\nabla'\rho_1\right)/n\\
\nonumber
&+&\frac{1}{2}\nabla\Re\left((\nabla\gamma)^2-(\nabla'\gamma)^2\right)
+\frac{1}{2}\nabla\left(\nabla^2 -\nabla'^2\right)\beta\\
&-&\frac{1}{2n(\br t)}\int d^3r'
\Delta P(\br\br' t)\nabla v\ee(|\br'-\br|)
\label{eq:potential}
\eea
where in the first three lines we have omitted the arguments and it is
understood that $\br'$ and $\br$ are set equal after all the
derivatives are taken. ${\bf B}(\br t)$ is an undetermined vector
whose role, together with an additional constant $C(t)$, is to ensure
satisfaction of a boundary condition on the potential.

Now the prescription is to pick an initial
state which has the same initial density and initial first
time-derivative of the density as the state $\Psi$; that is, require
$\gamma(\br\br 0)=0$ and
Eq. (\ref{eq:divngradalpdiag}) taken at $t=0$. Then one
can evaluate Eq. (\ref{eq:potential}) at $t=0$
and so find $v({\bf r} 0)$.
The procedure for $t>0$ is described in detail in Ref. \cite{L99}.
In order for this procedure to yield a well-behaved physical
potential, one needs
to first check that the initial potential 
is not divergent at
infinity. (Equivalently, we may require that the elements of the
momentum-stress tensor appearing in Eq. (\ref{eq:SL}) do not diverge at
infinity).  This gives an additional restriction on the initial
state.
In the one-electron case, this restriction rules out {\it any} other
candidate for an initial wavefunction which evolves with the same
density as another wavefunction does in another potential: there is
no way to pick ${\bf B}(\br 0)$ or the constant $C(0)$ to satisfy
any physical boundary condition discussed above.  In the
many-electron case, our additional condition restricts the allowable
wavefunctions, but does not render the question of initial-state
dependence moot as in the one-electron case.

\section{Examples}
\label{s:examples}
\subsection{One electron in one dimension}
\label{s:one-e in one-d}
By studying the \td Schr\"odinger equation for one electron in one
dimension, it is simple to find explicitly the potential $\tilde v(x t)$ which
cajoles $\tilde\phi(x t)$ into evolving with the same density as $\phi(x t)$
which evolves in a different potential $v(x t)$.  Consistent with
the conclusions above, this potential diverges to $-\infty$ at large
$x$, which is unphysical. The initial state is pathological in the
sense that its expectation value of momentum, kinetic energy and
potential energy all diverge. A phase-space picture helps us to see
how such a potential can hold a localized density.

Inserting Eq. (\ref{eq:psi}) into the \td Schr\"odinger Eq. (\ref{eq:TDSE})
and calculating the derivatives, we obtain
\ben
\Delta v=
i\alpha''/2 +i \alpha' \phi'/\phi -
{\alpha'}^2/2 -\dot\alpha  = 0. 
\label{eq:tdsepsi}
\een
where primes denote spatial derivatives.
We now write $\phi$ in terms of an amplitude and phase
\ben
\phi(x t) = \sqrt{n(x t)}\ \exp\left(i\theta(x t)\right).
\een
Substituting into Eq. (\ref{eq:tdsepsi})
and setting the real and imaginary terms separately to zero, 
yields
\ben
\Delta v= -\dot\alpha -\alpha'\theta'-{\alpha'}^2/2;
~~~~~~
\alpha''+\alpha' \n'/\n =0.
\label{eq:repart}
\een
We find for $\alpha'$:
\ben 
\alpha' = c(t)/n(x t).
\label{eq:dalpdx}
\een 
We observe that this is also obtained when
Eq. (\ref{eq:divngradalpha}) (which arose from setting the time-derivatives
of the densities to be equal) is considered in one-dimension.
Integrating once more gives
 \ben
\alpha(x t) =c(t)\int^x \frac{dx'}{n(x' t)} +d(t)
\label{eq:alpha}
\een
Plugging this solution into Eq. (\ref{eq:repart}) gives
\bea
\nonumber
\Delta v(x t)& =& -\frac{c(t)\theta'(x t)}{n(x t)}
-\frac{1}{2}\frac{c^2(t)}{n^2(x t)} \\
&-&\dot c(t) \int^x \frac{dx'}{n(x' t)}
+c(t)\int^x \frac{\dot n(x' t)}{n^2(x' t)}dx'
- \dot d(t)
\label{eq:pot1e1d}
\eea
We see immediately the divergence of this potential (for non-zero
$c(t)$) where the density decays at large $x$.
This demonstrates by explicit solution of the \td Schr\"odinger
equation that the only potentials in which a density can be made to
evolve as the density in some other potential are unphysical,
consistent with the conclusions of section \ref{s:one}.

The state $\tilde\phi $ oscillates more and more wildly as $x$ gets
larger in the tails of the wavefunction.  Although the decay of the
density at large distances unweights the rapidly oscillating phase, it
is not enough to cancel  the infinite energy that the wild
oscillations contribute. Calculating the expectation value of momentum
or kinetic energy in the state Eq. (\ref{eq:psi}) with $\alpha$ given by
Eq. (\ref{eq:alpha}),
for a typical density and state $\phi$ ({\it e.g.}, one which decays
exponentially at large $x$), we find they blow up. 

\begin{figure}
\centerline{
\psfig{figure=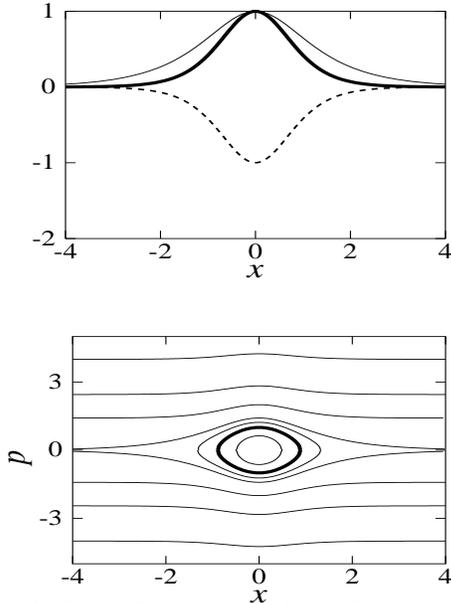,width=6cm,height=8cm}}
\caption{The lower figure shows classical phase-space contours for the ${\rm sech^2}$ well. The top figure shows the potential (dashed line), the wavefunction (solid line) corresponding to the heavily drawn contour in the phase-space below and its density (thick solid line). }
\label{fig:eckplot}
\end{figure}

\begin{figure}
\centerline{
\psfig{figure=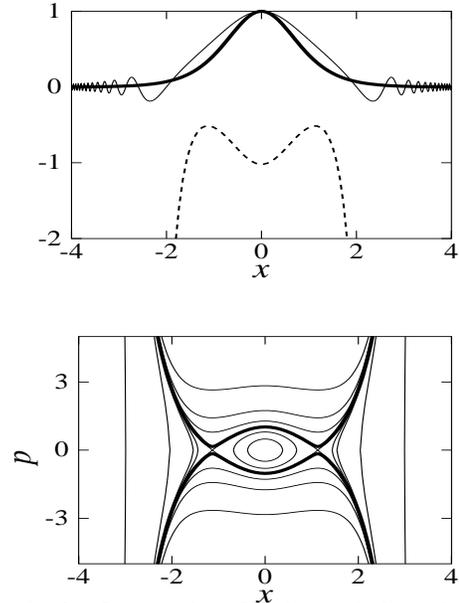,width=6cm,height=8cm}}
\caption{As for figure \ref{fig:eckplot} but for the pathological potential described in the text with $c=0.2$. The thinner solid line in the top figure is the real part of $\tilde \phi$. }
\label{fig:eckplotdiv}
\end{figure}

It may be at first glance striking that a potential which plunges to
minus infinity at large distances can hold a wavefunction which is
localized in a finite region in space. Consider the special case in
that $\phi(x t)$ is an eigenstate of a time-independent potential
$v(x)$. Let us also choose $c(t) = c$ to be time-independent, so that
$\tilde v(x)$ is also time-independent and $\tilde\phi(x t)$ is an
eigenstate of it. Let the density $n(x) = \vert \phi(x t)\vert^2
=\vert \tilde\phi(x t)\vert^2$ be localized at the origin. For
example, $v$ might be a potential well with flat asymptotes.  Then we
have the interesting situation where the eigenstate $\tilde\phi(x)$ is
localized at the origin of its potential $\tilde v$ which plummets to
$-\infty$ at large $x$. In Figs. \ref{fig:eckplot} and
\ref{fig:eckplotdiv} we have plotted the potentials, the densities and
the (real part of the) wavefunctions for the two cases; notice the
steep cliffs of $\tilde v$ and the rapid oscillations of $\tilde\phi$
as $x$ gets large as we predicted.  We chose the potential $v = -{\rm
sech}^2(x)$, and the state $\phi$ as its ground-state. In the lower
half of each figure are the classical phase space pictures (the
classical energy contours) for the two potentials; to a good
approximation the quantum eigenstates lie on those contours which have
the correctly quantized energy (semiclassical approximation
\cite{H77}). In Fig. \ref{fig:eckplot} is the situation for potential
$v$; $\phi$ lies on the heavily drawn contour shown, which is a bound
state oscillating inside the well. In Fig. \ref{fig:eckplotdiv} is
the situation for $\tilde v$; the heavy contour that $\tilde\phi$ lies
on is of a different nature, not bound in any region in space. However
its two branches fall away from the origin very sharply, so that
although they eventually extend out to large $x$, the projection on
the $x-$ plane is much denser near the origin than further away. This
is how such a potential can support a localized density near the
origin. The phase of the wavefunction in a semiclassical view is given
by the action integral $\int p(x) dx$ along the contour and the large
step in $p$ that is made in a short step in $x$ thus implies that the
phase oscillates rapidly.  The tails of the density, which is the same
in both cases, arise from fundamentally different processes: in the
case of the simple ${\rm sech^2}$ well (Fig. \ref{fig:eckplot}) the
tails arise from classically forbidden tunneling whereas in the case
of the divergent potential (Fig. \ref{fig:eckplotdiv}) they are
classically allowed but have exponentially small amplitude.


Finally, we relate the result for $\tilde v$ (Eq. (\ref{eq:pot1e1d}))
to that obtained from the
approach in \cite{L99} (and outlined in \ref{s:many}).  Observe that
for one electron, the initial conditions on $\tilde\phi$ required in
\cite{L99} are the same as our 
Eqs. (\ref{eq:psi}) and (\ref{eq:divngradalpha}) 
(or, equivalently Eq. (\ref{eq:dalpdx})),
each evaluated at $t=0$. In one-dimension it
is then a straightforward exercise to calculate the terms of Eq.
(\ref{eq:potential}) at $t=0$, and, if we disregard the boundary condition,
we can thus obtain an equation for the slope of the potential. 
This
potential gradient is consistent with the equation (\ref{eq:pot1e1d})
evaluated at $t=0$ that we obtained by the \td Schr\"odinger's
equation above: we get here
\ben
\Delta v'= (c^2 n'+2 c n\dot n+2c j + n^2 f)/n^3
\label{eq:potrvl1e1d}
\een
where $n=n(x 0), j=j(x 0), v=v(x 0), c=c(0)$, and $f$ is a constant
to be determined by the boundary condition, $\Delta v \to 0$
at $\infty$.
However in the
present one-electron case such a boundary condition cannot be
satisfied.

\subsection{Two non-interacting electrons in one-dimension}
\label{s:two-e in one-d}

Let $\phi_1$ and $\phi_2$
represent the initial orbitals for an initial state $\Phi$ and
$\tilde \phi_1$ and $\tilde \phi_2$ represent those for
another initial state $\widetilde \Phi$. 
We choose 
\ben
\tilde \phi_i(x) = \phi_i(x)\ \exp{i\theta_i(x)}
\label{eq:orbitals}
\een
where the $\theta_i (x)$ are real functions.
This form guarantees the densities are the same initially. 
The difference of initial 
current between $\widetilde \Phi$ and $\Phi$ is
\ben
\Delta j =\theta_1'(x)\vert \phi_1(x) \vert^2
+ \theta_2'(x)\vert \phi_2(x) \vert^2,
\label{eq:deljorb}
\een
where the prime indicates differentiation, 
and so the condition of equal initial $\dot n$ becomes
\ben
\frac{\partial}{\partial x} \left(\theta_1'(x) \vert \phi_1(x) \vert^2 + \theta_2'(x)\vert \phi_2(x) \vert^2\right) = 0.
\een
The choice
\ben
\theta_1'(x) = c\vert \phi_2(x)\vert^2, \,
{\rm  and}  \,\, \theta_2'(x) = -c\vert \phi_1(x)\vert^2.
\label{eq:f1f2}
\een
where $c$ is some constant, ensures eq. \ref{eq:deljorb} is satisfied.

To simplify the calculation of the potential gradient
in Eq. (\ref{eq:potential}) further,
we take the orbitals $\phi_1$ and $\phi_2$ to be
real and take the density to be time-independent. 
After straightforward  calculations we arrive at
\bea
\Delta v'=-c^2(n'\phi_1^2 \phi_2^2/n +2\phi_1\phi_2(\phi_2 \phi_1'+\phi_1 \phi_2'))
\label{eq:finalpot2e}
\eea
This gives the initial gradient of the potential in which  $\widetilde\Phi$ 
will
evolve with the same density as that of $\Phi$
at $t=0$. 

These equation was used to make Fig. \ref{fig:eg2}, for which $c=4$.
The orbitals $\phi_i(x)$ are just the lowest and first
excited states of the harmonic oscillator of force constant $k=1$.
Although this kind of potential is not strictly
allowed because it does not remain finite at $\infty$, we expect that
the results still hold for well-behaved initial states: 
the difference between the potentials for $\Phi$ and
$\widetilde\Phi$ vanishes at $\infty$. Moreover, it, or
rather, the interacting 3-D version (Hooke's atom) is instructive for
studying properties of density functionals (see e.g. \cite{HPB99})
because an exact solution is known.
We can make the dips on the right of the figure arbitrarily big,
simply by increasing $c$.
Note that  the alternative orbitals  probably do not yield
 an eigenstate of this potential. 
In the next instant, this alternative potential will change,
in order to keep the density constant.  This change can be calculated
using Van Leeuwen's prescription.

\section{Acknowledgments}
We thank Robert van Leeuwen for entertaining discussions.
This work was supported by NSF grant number CHE-9875091,
and KB was partially supported by the Petroleum Research Fund.
Some of the original ideas were discussed at the Aspen center for
physics.

\end{document}